\documentclass[sigconf,9pt]{acmart}

\usepackage{geometry}
\usepackage{listings}
\usepackage{algorithm}
\usepackage{hhline}
\usepackage{colortbl}
\usepackage{centernot}
\usepackage{pbox}
\usepackage{amsmath}
\usepackage{amsfonts}
\usepackage{url}
\usepackage{bm}
\usepackage{comment}
\usepackage{graphicx}
\usepackage{subfigure}
\usepackage{epstopdf}
\usepackage{multirow}
\usepackage{color}
\usepackage{tabu}
\usepackage{mdframed}
\usepackage{syntax}
\usepackage{fancyvrb}
\usepackage{multicol}
\usepackage{booktabs,dcolumn,caption}
\usepackage[noend]{algpseudocode}
\usepackage{algpseudocode}
\usepackage{paralist}
\usepackage{stmaryrd}
\usepackage{tikz}
\usepackage[referable]{threeparttablex}
\usepackage{tabularx}
\usepackage{booktabs}
\usepackage{array}
\usepackage{fancyhdr}
\usepackage{float}
\usepackage{xspace}
\usepackage{pifont}
\usepackage{balance}
\usepackage{xcolor}

\definecolor{codegreen}{rgb}{0,0.6,0}
\definecolor{codegray}{rgb}{0.5,0.5,0.5}
\definecolor{codepurple}{rgb}{0.58,0,0.82}
\definecolor{backcolour}{rgb}{0.95,0.95,0.92}
\lstdefinestyle{mystyle}{
    backgroundcolor=\color{backcolour},   
    commentstyle=\color{codegreen},
    keywordstyle=\color{magenta},
    numberstyle=\tiny\color{codegray},
    stringstyle=\color{codepurple},
    basicstyle=\ttfamily\footnotesize,
    breakatwhitespace=false,         
    breaklines=true,                 
    captionpos=b,                    
    keepspaces=true,                 
    numbers=left,                    
    numbersep=5pt,                  
    showspaces=false,                
    showstringspaces=false,
    showtabs=false,                  
    tabsize=2
}
\newcommand*\circled[1]{\tikz[baseline=(char.base)]{
            \node[shape=circle,fill,inner sep=1pt,scale=0.8] (char) {\textcolor{white}{#1}};}}
\usepackage{listings}
\usepackage{blindtext}
\usepackage{tcolorbox}
\usepackage{graphicx}
\usepackage{cleveref}

\definecolor{codegreen}{rgb}{0,0.6,0}
\definecolor{codegray}{rgb}{0.5,0.5,0.5}
\definecolor{codepurple}{rgb}{0.58,0,0.82}
\definecolor{backcolour}{rgb}{0.95,0.95,0.92}

\usepackage{pifont}
\usepackage{tablefootnote}

\usepackage{graphicx}
\usepackage{adjustbox}

\definecolor{princetonorange}{RGB}{255,143,0}

\definecolor{lightgreen}{RGB}{198, 224, 183}

\definecolor{lightred}{RGB}{240, 205, 176}

\newcommand{\question}[1]{\textcolor{orange}{#1}}

\usepackage{colortbl}
\usepackage{dcolumn}

\definecolor{greenDeep}{RGB}{0,170,0}
\definecolor{greenSlightDeep}{RGB}{0,205,0}
\definecolor{greenShallow}{RGB}{0,255,0}
\definecolor{greenShallower}{RGB}{160,255,0}
\definecolor{orangeShallow}{RGB}{255,190,0}
\definecolor{orangeDeep}{RGB}{255,80,0}
\definecolor{orangeDeeper}{RGB}{255,40,0}
\definecolor{redDeep}{RGB}{255,0,0}

\definecolor{redLight}{RGB}{255,128,114}

\def\zz#1{%
\ifdim#1pt>4.9pt\cellcolor{greenDeep}\else
\ifdim#1pt>3.9pt\cellcolor{greenSlightDeep}\else
\ifdim#1pt>2.9pt\cellcolor{greenShallower}\else
\ifdim#1pt>2.9pt\cellcolor{yellow}\else
\ifdim#1pt>1.9pt\cellcolor{orangeShallow}\else
\ifdim#1pt>1.9pt\cellcolor{orange}\else
\ifdim#1pt>0.9pt\cellcolor{orange}\else
\ifdim#1pt>0.9pt\cellcolor{orangeDeep}\else
\cellcolor{orangeDeep}\fi\fi\fi\fi\fi\fi\fi\fi
#1}

\copyrightyear{2024} 
\acmYear{2024} 
\setcopyright{acmlicensed}\acmConference[ICCAD '24]{Proceedings of the ACM/IEEE International Conference on Computer-Aided Design}{October 27--31, 2024}{New Jersey, NY, USA}
\acmBooktitle{Proceedings of the ACM/IEEE International Conference on Computer-Aided Design (ICCAD '24), October 27--31, 2024, New Jersey, NY, USA}

\begin{document}

\title{OpenLLM-RTL: Open Dataset and Benchmark for LLM-Aided Design RTL Generation}


\subtitle{Invited Paper}

\author{Shang Liu$^*$, Yao Lu$^*$, Wenji Fang$^*$, Mengming Li, Zhiyao Xie\textsuperscript{$\dagger$}}

\affiliation{%
 \vspace{.08in}
 \institution{Hong Kong University of Science and Technology (HKUST)}
 \vspace{.04in}
 \institution{\{sliudx, yludf, wfang838, mengming.li\}@connect.ust.hk, \ \ eezhiyao@ust.hk}
 \country{}
 \vspace{.03in}
}




\begin{abstract}

The automated generation of design RTL based on large language model (LLM) and natural language instructions has demonstrated great potential in agile circuit design. However, the lack of datasets and benchmarks in the public domain prevents the development and fair evaluation of LLM solutions. This paper highlights our latest advances in open datasets and benchmarks from three perspectives: (1) RTLLM 2.0, an updated benchmark assessing LLM's capability in design RTL generation. The benchmark is augmented to 50 hand-crafted designs. Each design provides the design description, test cases, and a correct RTL code. (2) AssertEval, an open-source benchmark assessing the LLM’s assertion generation capabilities for RTL verification. The benchmark includes 18 designs, each providing specification, signal definition, and correct RTL code. (3) RTLCoder-Data, an extended open-source dataset with 80K instruction-code data samples. Moreover, we propose a new verification-based method to verify the functionality correctness of training data samples. Based on this technique, we further release a dataset with 7K verified high-quality samples. These three studies are integrated into one framework, providing off-the-shelf support for the development and evaluation of LLMs for RTL code generation and verification. Finally, extensive experiments indicate that LLM performance can be boosted by enlarging the training dataset, improving data quality, and improving the training scheme.

\end{abstract}

\maketitle
\pagestyle{plain}

\begingroup\renewcommand\thefootnote{$*$}
\footnotetext{Equal Contribution}
\endgroup

\begingroup\renewcommand\thefootnote{$\dagger$}
\footnotetext{Corresponding Author}
\endgroup

\section{Introduction}\label{sec:intro}

In recent years, large language models (LLMs) such as GPT~\cite{gpt4} have demonstrated remarkable performance in natural language processing (NLP). Inspired by this progress, researchers have started exploring the adoption of LLMs in agile hardware design~\cite{chen2024dawn}. A promising direction that attracts the most attention is automatically generating design RTL based on natural language instructions~\cite{yang2024formaleval, chang2023chipgpt, blocklove2023chip, lu2023rtllm, liu2023verilogeval, thakur2023benchmarking, thakur2023autochip, nair2023generating, liu2023chipnemo, pei2024betterv, nakkab2024rome, liu2023rtlcoder, zhao2024codev, zhang2024mg, goh2024english}. In modern VLSI design flow, design teams typically exert great effort to implement precise design functionality in design RTL using hardware description languages (HDLs). Now given design functionality descriptions in natural language (e.g., specification), LLM solutions target directly generating corresponding HDL code such as Verilog, VHDL, and Chisel from scratch. This LLM-based design generation technique can potentially revolutionize the existing HDL-based VLSI design process, relieving designers from the tedious HDL coding tasks.  Compared with well-explored \emph{predictive} machine learning (ML)-based solutions in EDA~\cite{rapp2021mlcad, xie2022intelligent}, such \emph{generative} methods may benefit the hardware design process more directly.

\begin{figure}[!t]
  \centering
  \includegraphics[width=1\linewidth]{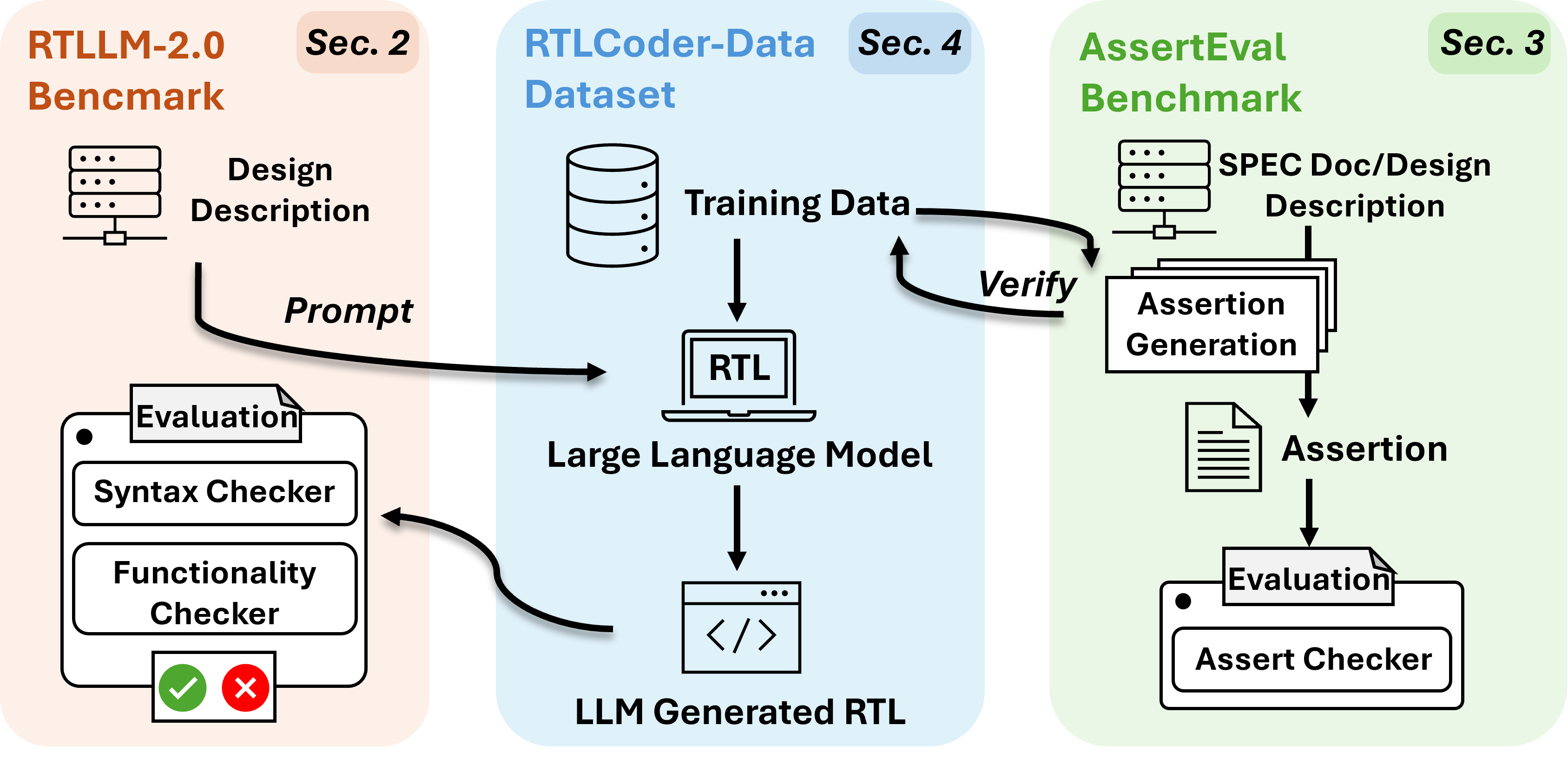}
  \vspace{-.2in}
  \caption{This paper presents open-source benchmarks and dataset for LLM-assisted RTL generation and verification.} 
  \label{fig:intro-fig}
  \vspace{-.1in} 
\end{figure}

In addition to the generation of RTL (i.e., HDL code) itself, the verification of RTL correctness is equally important and challenging in modern VLSI design.  Functional verification ensures the RTL implementation satisfies its specification. Assertion-based verification (ABV)~\cite{witharana2022survey} employs assertions derived from specifications to verify the functional behavior of RTL designs. ABV can be conducted through either simulation or formal property verification (FPV), with assertions often expressed using SystemVerilog Assertions (SVAs).  However, a major challenge in ABV is to obtain sufficient, high-quality assertions. Existing research on automating assertion generation includes dynamic assertion mining based on simulation traces~\cite{germiniani2022harm, danese2017team, vasudevan2010goldmine}, static generation using predefined design-specific templates~\cite{orenes2021autosva, fang2023r}, and the direct translation of natural language specifications into assertions~\cite{kande2024security, orenes2023using, sun2023towards, harris2016glast,krishnamurthy2019controlled,zhao2019automatic,krishnamurthy2019ease,frederiksen2020automated,keszocze2019chatbot,parthasarathy2021spectosva,aditi2022hybrid}. LLM solutions~\cite{fang2024assertllm, kande2024security, mali2024chiraag, orenes2023using,   sun2023towards,liu2024domain,bhandari2024llm} turn out to be also promising in generating assertions for design RTL verification.

Many existing works directly prompt commercial LLMs like GPT-3.5/GPT-4 for RTL code generation~\cite{chang2023chipgpt, blocklove2023chip, lu2023rtllm, thakur2023autochip, nair2023generating, nakkab2024rome} or verification~\cite{fang2024assertllm, huang2024towards, bhandari2024llm, xu2024meic, yang2024formaleval, tsai2023rtlfixer}, without proposing new datasets or models. However, reliance on commercial LLM tools limits in-depth research exploration and further model customization. More importantly, users of commercial LLM solutions unavoidably have data privacy concerns, since all instructions have to be uploaded to LLM providers like OpenAI. Such privacy concerns are especially critical in the IC design industry. In addition, commercial LLMs may not ensure reliable service with a low response latency.


To develop our own customized or open-source LLM solutions for RTL generation or verification, a primary challenge is the limited availability of circuit data. Unlike the huge amount of text and image resources in the public domain, circuit designs are the most important intellectual property (IP) of semiconductor companies, who typically strongly oppose sharing their designs. Such limited circuit data sharing is a long-standing issue not only for academia but also among different design teams within a single company. This data availability problem leads to a lack of datasets and benchmarks, preventing both the development and fair evaluation of LLM solutions in hardware design.

\begin{table}[!t]
      \centering
      \vspace{.1in}
     \hspace{-.1in}
     \setlength{\tabcolsep}{0.5em}
      \renewcommand{\arraystretch}{1.2}
      \resizebox{0.49\textwidth}{!}{
        \begin{tabular}{ |c||c| } 
        \hline
        \multicolumn{2}{|c|}{LLM-Assisted RTL Generation}  \\ 
        \hline
        \hline
           Prompt Engineering  &   \cite{chang2023chipgpt, blocklove2023chip, lu2023rtllm, thakur2023autochip, nair2023generating, nakkab2024rome}\\  
         \hline
          \multirow{2}{*}{Closed Dataset}   &  VerilogEval~\cite{liu2023verilogeval}, BetterV~\cite{pei2024betterv}, ChipNemo~\cite{liu2023chipnemo} \\
             &   Chang et al.~\cite{chang2024data}, OriGen~\cite{cui2024origen}, CodeV~\cite{zhao2024codev}    \\
         \hline
          \multirow{2}{*}{\textbf{Open Benchmark}}   &   RTLLM~\cite{lu2023rtllm},  VerilogEval~\cite{liu2023verilogeval}, RTL-repo~\cite{allam2024rtl} \\ 
            &   \textbf{RTLLM 2.0 (Section~\ref{sec:rtllm})}  \\
         \hline
        \textbf{Open Dataset}     &   \multirow{2}{*}{Thakur et al.~\cite{thakur2023benchmarking},  Wang et al.~\cite{wang2024large}}    \\
            (code only)      &       \\
         \hline
         \textbf{Open Dataset}   & RTLCoder~\cite{liu2023rtlcoder}, MG-Verilog~\cite{zhang2024mg},  Goh et al.~\cite{goh2024english} \\
           (instruction-code)    &     \textbf{RTLCoder-Data (Section~\ref{sec:rtlcoder})}  \\
                 \hline
             \multicolumn{2}{c}{    }    \\[-.3em]
        \hline
        \multicolumn{2}{|c|}{LLM-Assisted RTL Verification \& Debugging}  \\ 
        \hline
          \hline
        Prompt Engineering  &   \cite{fang2024assertllm, bhandari2024llm, xu2024meic,kande2024security, mali2024chiraag, orenes2023using,   sun2023towards,liu2024domain,huang2024towards,tsai2023rtlfixer}\\
        \hline
         Closed Dataset   &    HDLdebugger~\cite{yao2024hdldebugger} \\
         \hline
          \textbf{Open Benchmark}  &    \textbf{AssertEval (Section~\ref{sec:asserteval})}  \\
          \hline 
        \end{tabular}
       }
       \vspace{.04in}
        \caption{Existing explorations in LLM-aided design RTL generation and verification, with a focus on works that adopt or propose new datasets and benchmarks.} 
               \vspace{-.3in}
        \label{priorWorks}
\end{table}

Table~\ref{priorWorks} summarizes existing efforts in LLM-assisted RTL generation and verification, with a focus on open-source datasets and benchmarks. Open benchmarks~\cite{lu2023rtllm, liu2023verilogeval, allam2024rtl} are vitally important for a fair evaluation of LLM solutions. In addition to prompting GPT, many works tried to construct their own LLMs with either open-source~\cite{liu2023rtlcoder, thakur2023autochip, wang2024large, goh2024english, zhang2024mg} or closed-source~\cite{liu2023verilogeval, pei2024betterv, liu2023chipnemo, chang2024data, cui2024origen, zhao2024codev} datasets. Among the open-sourced dataset, several of them~\cite{thakur2023autochip, wang2024large} only provide RTL code, without alignment with the RTL generation tasks based on natural language instructions. In comparison, some open datasets~\cite{liu2023rtlcoder, zhang2024mg, goh2024english} provide a pair of natural language instruction (i.e., LLM input) and code (i.e., expected LLM output) as one data sample. These datasets are better aligned with the RTL generation task and benefit the LLM fine-tuning process.

In this paper, as summarized in Table~\ref{priorWorks} and Figure~\ref{fig:intro-fig}, we highlight our latest advances in open datasets and benchmarks for LLM-assisted design and integrate them into a unified framework. It consists of three major components. 

\begin{enumerate}
    \item In Section~\ref{sec:rtllm}, we present an open-source \textbf{benchmark named RTLLM 2.0 for evaluating the performance of LLM-assisted RTL generation}\footnote{RTLLM 2.0 is in https://github.com/hkust-zhiyao/RTLLM.}. It provides 50 RTL designs. It is an extension of our proposed benchmark RTLLM~\cite{lu2023rtllm}, which originally provided 30 designs. For each design, we provide the functionality description, test cases, and correct RTL design handcrafted by human engineers. 
    \item In Section~\ref{sec:asserteval}, we present an open-source \textbf{benchmark named AssertEval for LLM-assisted RTL verification}\footnote{AssertEval is in https://github.com/hkust-zhiyao/AssertLLM.}. It provides 18 designs to evaluate the generation of assertions by LLMs. These designs cover a diverse spectrum of applications. For each design, we provide the specification document, golden RTL code, and the script for FPV. 
    \item In Section~\ref{sec:rtlcoder}, we present an open-source \textbf{dataset named RTLCoder-Data for training the LLM for RTL generation}\footnote{RTLCoder-Data (both 80K and 7K) is in https://github.com/hkust-zhiyao/RTL-Coder.}. This dataset provides 80K (thousand) samples, with each sample being a code generation instruction and corresponding RTL code. This is an extension of the dataset released in our proposed RTLCoder~\cite{liu2023rtlcoder}, which originally provided 27K samples. 
\end{enumerate}

In addition, a challenge in dataset generation is the difficulty in checking the correctness of data samples. RTLCoder~\cite{liu2023rtlcoder} has proposed both \emph{instruction checker} and \emph{code checker}, evaluating the diversity introduced by new instructions and the syntax correctness of new code, respectively. 
However, no data generation method can automatically check whether the code has the correct functionality (i.e., same functionality as described in the instruction). In this paper, we explore an innovative method to verify training data correctness by generating assertions for each sample. In this way, we further generate and release a \textbf{verified 7K-sample dataset for training LLM for RTL generation}, which is also introduced in Section~\ref{sec:rtlcoder}. Finally, we trained and compared various LLM solutions to study the factors that affect LLM performance in RTL generation.

\begin{table*}[!]
\begin{center}
\vspace{-.1in}
 \resizebox{\textwidth}{!}{
\begin{tabular}{p{3cm}l||p{2.5cm}l}
\toprule[1.2pt]
\multicolumn{2}{c||}{\textbf{Arithmetic Modules}} & \multicolumn{2}{c}{\textbf{Memory   Modules}} \\ \midrule
\textbf{Design}                    & \textbf{Description} & \textbf{Design}           & \textbf{Description}   \\ \midrule
adder\_8bit                        & An 8-bit adder                                                                                                       & asyn\_fifo               & An asynchronous FIFO 16×8 bits                                                                                             \\
adder\_16bit                       & A 16-bit adder implemented with full adders                                                                          & \textbf{LIFObuffer}      & A Last-In-First-Out buffer for temporary data storage                                                                      \\
adder\_32bit                       & A 32-bit carry-lookahead adder                                                                                       & right\_shifter           & Right shifter with 8-bit delay                                                                                             \\
adder\_pipe\_64bit                 & \begin{tabular}[c]{@{}l@{}}A 64-bit ripple carry adder based on 4-stage \\ pipeline\end{tabular}                     & \textbf{LFSR}            & \begin{tabular}[c]{@{}l@{}}A Linear Feedback Shift Register for generating \\ pseudo-random sequences\end{tabular}         \\
\textbf{adder\_bcd}                & A BCD adder for decimal arithmetic operations                                                                        & \textbf{barrel\_shifter} & A barrel shifter for rotating bits efficiently                                                                             \\ 
\textbf{sub\_64bit}                & A 64-bit subtractor for high-precision arithmetic                                                                    & RAM                      & 8x4 bits true dual-port RAM                              \\ 
\textbf{multi\_8bit}               & \begin{tabular}[c]{@{}l@{}}An 8-bit multiplier based on shifting and adding \\ operation\end{tabular}                &  \textbf{ROM}             & A Read-Only Memory module for storing fixed data          \\   \cmidrule{3-4} 
multi\_16bit                       & \begin{tabular}[c]{@{}l@{}}A 16-bit multiplier based on shifting and adding \\ operation\end{tabular}                & \multicolumn{2}{c}{\textbf{Miscellaneous Modules}}                               \\ \cmidrule{3-4} 
multi\_booth\_8bit                 & An 8-bit booth-4 multiplier                                                                                          &  \textbf{Design}           & \textbf{Description}                                                                          \\   \cmidrule{3-4} 
multi\_pipie\_4bit                 & A 4-bit unsigned number pipeline multiplier                                                                          & \textbf{clkgenerator}    & A clock generator for providing timing signals                                                                             \\
multi\_pipie\_8bit                 & An 8-bit unsigned number pipeline multiplier                                                                         & \textbf{instr\_reg}      & \begin{tabular}[c]{@{}l@{}}An instruction register module for holding and \\ processing CPU instructions\end{tabular}      \\
div\_16bit                         & A 16-bit divider based on subtraction operation                                                                      &  signal\_generator        & Generate various signal patterns                                                           \\
radix2\_div                        & An 8-bit radix-2 divider                                                                                             &  \textbf{square\_wave}    & A generator for producing square wave signals                                                                                  \\
\textbf{comparator\_3bit}          & A 3-bit comparator for comparing binary numbers                                                                      & alu                      & An ALU for 32bit MIPS-ISA CPU                                                                                              \\
\textbf{comparator\_4bit}          & A 4-bit comparator for comparing binary numbers                                                                      & pe                       & A Multiplying Accumulator for 32bit integer                                                                                \\
accu                               & Accumulates 8-bit data and output after 4 inputs                                                                     & freq\_div                & \begin{tabular}[c]{@{}l@{}}Frequency divider for 100M input clock, outputs \\ 50MHz, 10MHz, 1MHz\end{tabular}              \\
\textbf{fixed\_point\_adder}       & \begin{tabular}[c]{@{}l@{}}A fixed-point adder for arithmetic operations \\ with fixed precision\end{tabular}        & \textbf{freq\_divbyeven} & \begin{tabular}[c]{@{}l@{}}Frequency divider that divides input frequency by \\ even numbers\end{tabular}                  \\
\textbf{fixed\_point\_substractor} & \begin{tabular}[c]{@{}l@{}}A fixed-point subtractor for precise fixed-point \\ arithmetic\end{tabular}               & \textbf{freq\_divbyodd}  & \begin{tabular}[c]{@{}l@{}}Frequency divider that divides input frequency by odd \\ numbers\end{tabular}                   \\
\textbf{float\_multi}              & \begin{tabular}[c]{@{}l@{}}A floating-point multiplier for high-precision \\ calculations\end{tabular}               & \textbf{freq\_divbyfrac} & \begin{tabular}[c]{@{}l@{}}Frequency divider that divides input frequency by \\ fractional values\end{tabular}             \\\cmidrule(lr){1-2}
\multicolumn{2}{c||}{\textbf{Control   Modules}}                                                                                                            & calendar                 & Perpetual calendar with seconds, minutes, and hours                                                                        \\ \cmidrule(lr){1-2}
\textbf{Design}              & \textbf{Description} & traffic\_light           & \begin{tabular}[c]{@{}l@{}}Traffic light system with three colors and pedestrian \\ button\end{tabular}                    \\ \cmidrule(lr){1-2}
fsm                                & FSM detection circuit for specific input                                                                             & width\_8to16             & \begin{tabular}[c]{@{}l@{}}First 8-bit data placed in higher 8-bits of the 16-bit \\ output\end{tabular}                   \\
\textbf{sequence\_detector}        & Detect specific sequences in binary input                                                                            & synchronizer             & Multi-bit mux synchronizer                                                                                                 \\
counter\_12                        & Counter module counts from 0 to 12                                                                                   & edge\_detect             & Detect rising and falling edges of changing 1-bit signal                                                                   \\
JC\_counter                        & \begin{tabular}[c]{@{}l@{}}A 4-bit Johnson counter with specific cyclic \\ state sequence\end{tabular}               & pulse\_detect            & \begin{tabular}[c]{@{}l@{}}Extract pulse signal from the fast clock and create a \\ new one in the slow clock\end{tabular} \\
\textbf{ring\_counter}             & An 8-bit ring counter for cyclic state sequences                                                                     & parallel2serial          & Convert 4 input bits to 1 output bit                                                                                       \\
\textbf{up\_down\_counter}         & \begin{tabular}[c]{@{}l@{}}A 16-bit counter that can increment or \\ decrement based on control signals\end{tabular} & serial2parallel          & 1-bit serial input and output   data after receiving 6 inputs            \\\bottomrule[1.2pt]

\end{tabular}
}
\vspace{.05in}
\caption{RTLLM-2.0 benchmark description. The benchmark includes 50 designs across various applications, with \textbf{bold} designs representing newly added designs relative to RTLLM.} 
\label{benchmark2}
\vspace{-.15in}
\end{center}
\end{table*}

\section{RTLLM 2.0: Open Benchmark for RTL Generation}
\label{sec:rtllm}

\subsection{Overview of RTLLM 2.0}

 Our previously proposed RTLLM~\cite{lu2023rtllm} is a comprehensive open-source benchmark for design RTL generation with natural language. It supports the evaluation of any generated HDL format, including Verilog, VHDL, and Chisel, as long as it supports logic synthesis and RTL simulation. 
 
 RTLLM~\cite{lu2023rtllm} consists of 30 designs with a wide coverage of design complexities and scales. In RTLLM-2.0, we have expanded this collection to include 50 designs, of which the ones highlighted in \textbf{bold} in Table \ref{benchmark2} are newly added. This enlargement allows for a more thorough evaluation of different design types and sizes, offering a more comprehensive understanding of how RTL code generation performs across a wider variety of benchmarks. By increasing the number of designs, we can now explore a broader range of scenarios, better understand design intricacies, and assess performance more effectively in the context of RTL code generation. 
 
 Unlike RTLLM, which broadly classifies designs into \textit{Arithmetic} and \textit{Logic} types, RTLLM-2.0 takes a closer look by categorizing designs based on their specific functions and applications. By breaking down designs in RTLLM-2.0 according to their unique purposes, these detailed categories allow for better comparisons of how different models perform across various design types.

\subsection{Detailed Inspection of the Benchmark}
The benchmark RTLLM-2.0 dataset is meticulously categorized into four primary module classes: Arithmetic Modules, Memory Modules, Control Modules, and Miscellaneous Modules. Each class encompasses a variety of functional units pertinent to diverse computational and control tasks, as delineated in Table \ref{benchmark2}. This structured classification facilitates a comprehensive analysis and application of the dataset across multiple domains in digital system design.

\begin{table*}[!]
\begin{center}
\vspace{-.1in}
 \resizebox{1\textwidth}{!}{
\begin{tabular}{c|ccc|ccc|ccc|ccc|ccc}
\toprule
\textbf{Design Type}                                                                                                     & \multicolumn{3}{c|}{\textbf{Cryptographic   Unit}} & \multicolumn{3}{c|}{\textbf{Processor Core}} & \multicolumn{3}{c|}{\textbf{Arithmetic Unit}} & \multicolumn{3}{c|}{\textbf{Communication Protocol}} & \multicolumn{3}{c}{\textbf{Memory Controller}} \\ \hline \hline
\multirow{4}{*}{\begin{tabular}[c]{@{}c@{}}\textbf{Design   Name/}\\      \textbf{\# Page/}\\      \textbf{\# Signal}\end{tabular}} & AES              & 15        & 11        & amber      & 26     & 14     & ecg               & 9      & 12     & ethernet         & 42         & 54         & hpdmc           & 8        & 4        \\
                                                                                                         & sha3             & 17        & 9         & lxp32      & 59     & 22     & mac               & 24     & 34     & i2c              & 15         & 24         & sdc             & 26       & 53       \\
                                                                                                         & tiny\_aes        & 17        & 4         & minsoc     & 22     & 14     & pairing           & 13     & 8      & sockit           & 29         & 15         & sdr\_ctrl       & 28       & 46       \\
                                                                                                         &                  &           &           &            &        &        & tiny\_pairing     & 17     & 10     & uart             & 10         & 11         &                 &          &       \\ \bottomrule  
\end{tabular}

}
\end{center}
\caption{AssertEval benchmark description. The benchmark includes 18 open-source designs across various applications. For each design’s specification document, we list the number of file pages and the number of architectural signals under verification.}
\label{tbl:asserteval}
\vspace{-.2in}
\end{table*}

\textbf{Arithmetic Modules} comprise various adders, subtractors, multipliers, dividers, comparators, accumulators, and other specialized units like fixed-point arithmetic components. For instance, the adder subcategory includes 8-bit, 16-bit, 32-bit, and 64-bit pipelined adders, along with a BCD adder, addressing both general and specific arithmetic operations. Similarly, subtractors, multipliers, and comparators are available in different bit widths and configurations, such as a 64-bit subtractor, an 8-bit multiplier, and both 3-bit and 4-bit comparators, respectively.

\textbf{Memory Modules} are designed to handle data storage and retrieval with FIFO (First-In, First-Out) and LIFO (Last-In, First-Out) buffers, alongside various shifters including right shifters, LFSRs (Linear Feedback Shift Registers), barrel shifters, as well as RAM and ROM. The inclusion of asynchronous FIFO and LIFO buffers highlights the benchmark's capability to manage different memory access patterns efficiently.

\textbf{Control Modules} focus on state management and counting mechanisms, featuring finite state machines (FSMs), sequence detectors, and various counters such as a 12-bit counter, Johnson counter (JC\_counter), ring counter, and an up/down counter. These modules are crucial for controlling the flow of operations and ensuring sequential logic execution within digital systems.

\textbf{Miscellaneous Modules} cover a broad spectrum of functionalities including signal generation, RISC-V components, frequency dividers, and other essential units. Signal generation features modules like a signal generator and a square wave generator. The RISC-V category includes clock generators, instruction registers, ALU, and processing elements, essential for constructing RISC-V-based architectures. Frequency dividers are detailed with modules that divide by even, odd, and fractional values. Additionally, there are modules for specific applications such as calendars, traffic lights, data width converters, synchronizers, and various signal detection and conversion units.

This classification framework facilitates a nuanced understanding of the benchmark, highlighting its versatility and applicability across different domains of digital system design and analysis.

\subsection{Benchmark Evaluation}

\subsubsection{Overview of Test Files}

For each design, RTLLM-2.0 provides the following information in three separate files.

\textbf{Description} (\emph{design\_description.txt}): A natural language description of the target design's functionality, serving as a prompt for LLMs to generate RTL code. It includes the module name and all input/output (I/O) signals with names and widths, enabling automatic functionality verification with the provided testbench.  

\textbf{Testbench} (\emph{testbench.v}): A testbench containing multiple test cases with input and expected output values. It corresponds to the module name and I/O signals in  \emph{design\_description.txt} and is used to verify design functionality. 

\textbf{Correct Design} (\emph{designer\_RTL.v}): A reference design Verilog hand-crafted by human designers. By comparing with this reference design, we can quantitatively evaluate the design qualities of the automatically generated design. 

\subsubsection{Evaluation Metrics}

To systematically evaluate the generated design RTL, we summarize three progressive goals, which can all be evaluated with our benchmark. The first and basic goal is the \textbf{syntax goal}. It means the syntax of the generated RTL design should at least be correct. It can be verified by checking whether the design can be correctly synthesized into netlist by synthesis tools~\cite{design-compilier}. The second is \textbf{functionality goal}. It requires the generated RTL design to function as expected, verified by passing all test cases provided in \emph{testbench.v}. While the testbench samples a reasonable number of cases, passing them doesn't guarantee 100\% functionality correctness. If the design is correct in both syntax and functionality, it is considered successful. However, for practical use, its design qualities, including performance, power, and area (PPA), should also be desirable. This is the \textbf{quality goal}, verified by measuring PPA values after synthesis and layout.

\section{AsserEval: Open Framework and Benchmark for RTL Verification}

\label{sec:asserteval}

\subsection{Assertions Generation and Evaluation Framework}

Inspired by the potential of LLMs, translating natural language specifications into assertions has gained significant attention. Some works\cite{kande2024security, mali2024chiraag} leverage LLMs to convert human-extracted or human-written specification sentences into corresponding assertions. Other approaches, like AssertLLM~\cite{fang2024assertllm}, process entire specification documents directly, using LLMs to automatically extract assertion-related information from highly unstructured, multi-modal data, including descriptive text and waveform diagrams.

Despite the growing interest in LLM-based assertion generation, a universal evaluation method and benchmark are still unavailable. To address this challenge, we propose AssertEval, a benchmark and framework designed to evaluate the quality of LLM-based assertion generation across various VLSI designs.

\begin{figure}[!t]
  \centering
  \includegraphics[width=0.95\linewidth]{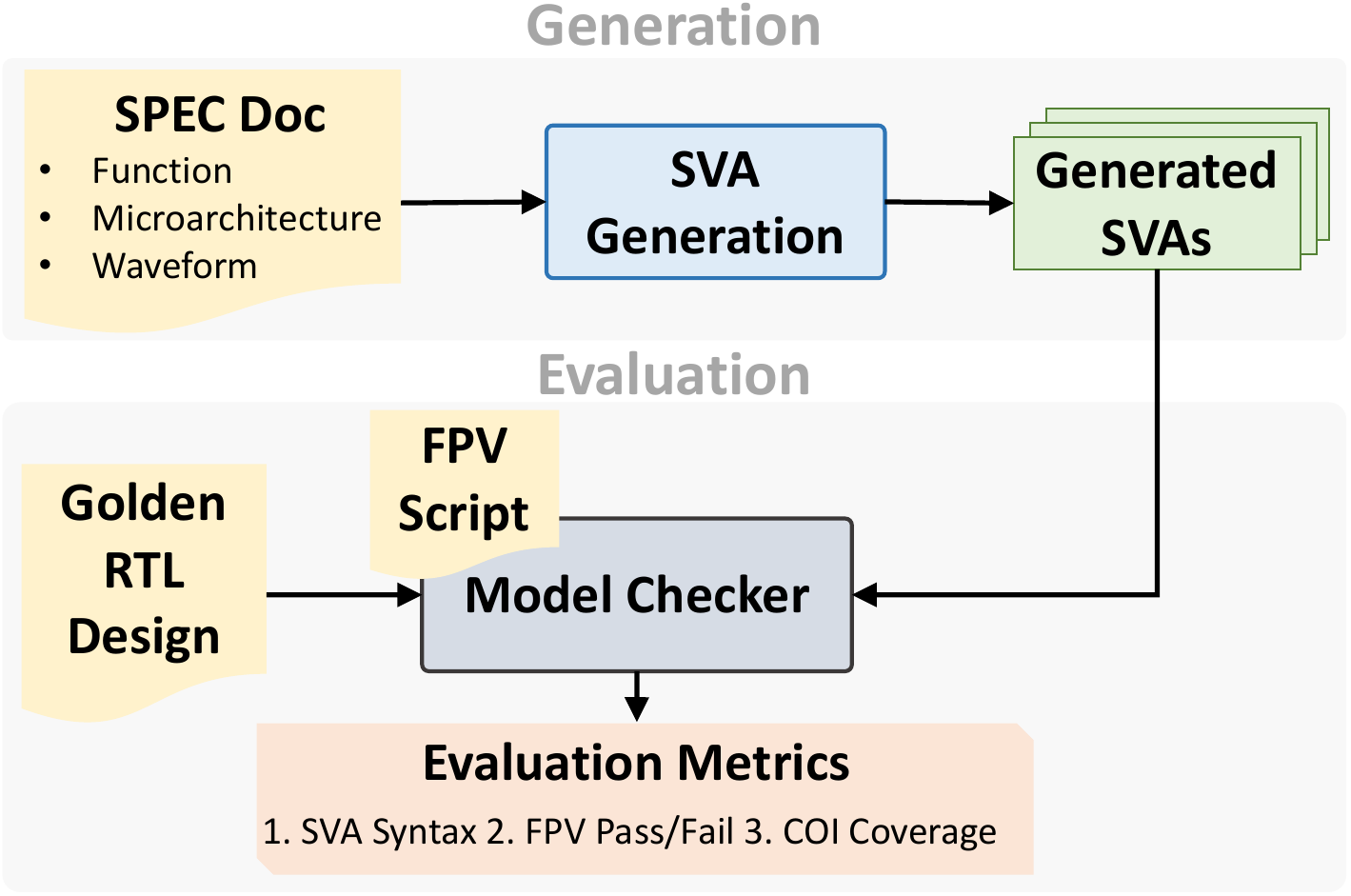}
  \caption{Evaluation of generated assertions using our benchmark. We provide natural language specification documents as input for the assertion generation process. The generated assertions are then evaluated against the provided golden RTL designs using the FPV technique. Three key metrics are employed to assess the quality of the generated assertions.}
  \label{fig:asserteval}
  \vspace{-.05in}
\end{figure}

The generation and evaluation flow is demonstrated in~\Cref{fig:asserteval}. For the generation process, we provide entire specification documents as input, users then generate assertions for each architectural signal with their own assertion generation methods. These assertions are then evaluated against our provided golden RTL implementations using formal property verification techniques.

After the assertion generation process, the framework automatically evaluates the quality of the generated assertions. Bug-free golden RTL implementations are provided for this evaluation. Based on these golden RTL designs, the generated assertions are verified through formal property verification (FPV) techniques.
After performing FPV, the following metrics are computed to evaluate the quality of generated assertions:
\begin{itemize}
    \item \textbf{Syntax}: checks if generated assertions have syntax errors.
    \item \textbf{FPV pass/fail}: when RTL designs are bug-free, an assertion that passes the FPV check is considered semantically correct, and conversely, a failure indicates an incorrect assertion.
    \item \textbf{COI coverage}: cone of influence (COI) coverage measures the percentage of design logic that is structurally connected to the properties. It is a common metric to evaluate the quality and usefulness of the generated properties.
\end{itemize}

\subsection{AssertEval Benchmark Description}

The benchmark AssertEval consists of 18 open-source designs that cover a diverse array of applications, including cryptographic units, processor cores, arithmetic units, communication protocols, and memory controllers. Considering the capability of existing LLM-based generation methods, we have collected specification documents that are fewer than 60 pages and contain fewer than 60 architecture-level signals. We list the detailed statistics for each design in~\Cref{tbl:asserteval}. Additionally, we provide an FPV script for Cadence JasperGold~\cite{Jasper}, which can be executed with a single button click for ease of use. For each design within the benchmark, our benchmark provides the following three files:

\begin{itemize}
    \item \textbf{Specification document}: This file contains the natural language specification for the design, providing a detailed description of the design architecture and functionality.\looseness=-1
    \item \textbf{Golden RTL implementation}: This file comprises the RTL design implementations that are strictly implemented according to the specification. The designs are verified to ensure it is free from bugs, serving as a reliable standard for evaluating the correctness of generated assertions.
    \item \textbf{FPV script}: This script automatically executes FPV, allowing users to execute the verification with a single click.\looseness=-1
\end{itemize}

The specification document is highly unstructured, with assertion-related information dispersed across various sections. Additionally, it includes multi-modal data (e.g., descriptive text and waveform diagrams), making the extraction of relevant details challenging.

Our provided specification typically includes seven key sections: 1) Summary: outlines the design's concepts and features; 2) IO ports: provides detailed information for the interface; 3) Registers: describes all the architecture-level registers in the design; 4) Operation: explains the operational procedures for dataflow and control; 5) Architecture: the high-level workflow and dataflow of the design; and 6) Usage examples: offers basic usage scenarios for the design.  For signals, the specification may only define critical architecture-level IO ports and registers, leaving the designers to detail internal signals for RTL implementations. 7) Waveform diagram: describe behaviors for different signals.

\begin{figure*}[!t]
\centering
\includegraphics[width=\textwidth]{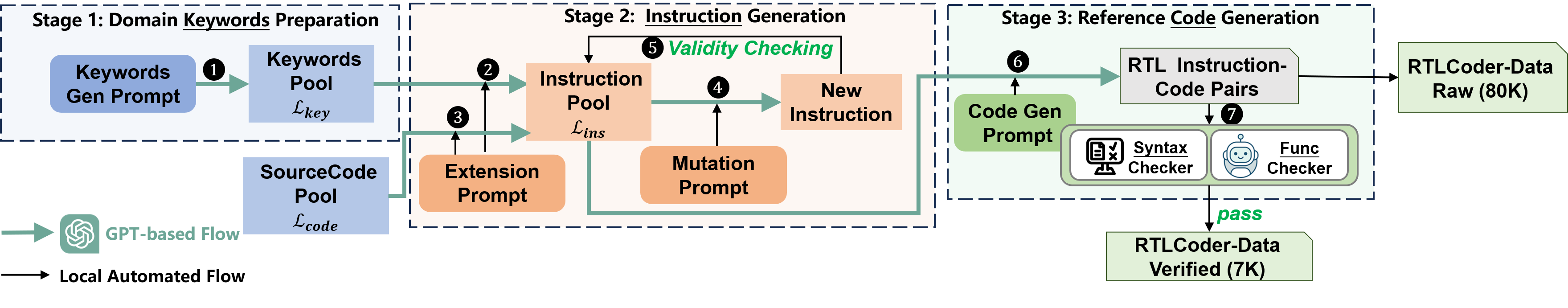}
\vspace{-.25in}
\caption{The automated training dataset generation flow to generate RTLCoder-Data. The framework is based on prior RTLCoder~\cite{liu2023rtlcoder}, but we proposed an innovative automated functionality checking method in Stage 3.} 
\label{fig:flow}
\end{figure*}

\section{Open Dataset for RTL Generation}

\label{sec:rtlcoder}

In this Section, we present an open-source dataset named RTLCoder-Data for training the LLM for RTL generation. It provides a large `raw dataset' with 80K samples, tripling the one previously released in RTLCoder~\cite{liu2023rtlcoder}. Moreover, we propose an innovative verification-based method to check the functionality correctness of each instruction-code data sample. Applying both the functionality checker and syntax checker, we further generate and release a high-quality verified dataset with 7K `mostly-correct' samples.

\subsection{Basic Dataset Generation Flow}

Our prior work RTLCoder~\cite{liu2023rtlcoder} has proposed an automated training dataset generation flow and generated 27K training samples, with each sample being a pair of design instruction (i.e., model input) and the reference RTL code (i.e., expected model output). The instruction can be viewed as the input question for LLMs, describing the desired circuit functionality in natural language. The reference code is the expected Verilog code that implements the functionality. This flow takes advantage of the powerful general text generation ability of the commercial tool GPT with several prompt templates. As Figure~\ref{fig:flow} shows, the flow includes three stages, which are summarized below.  


\textbf{Stage 1: Keywords Preparation.} The first stage of the data generation flow targets preparing RTL domain keywords for subsequent stages. At process \circled{1} in Figure~\ref{fig:flow} shows, GPT is requested to generate keywords related to digital IC design (i.e., commonly used logic components) based on a set of prompts. We obtain a keyword pool $\mathcal{L}_{key}$ with hundreds of digital design keywords.

\textbf{Stage 2: Instruction Generation.} The second stage targets generating sufficient instructions based on the initial keywords and Verilog source code. At process \circled{2}, existing keywords are extended from $\mathcal{L}_{key}$ to complete instructions. 
In addition to keyword-based instruction generation, we also generate instructions based on existing source code collected by us, as shown in process \circled{3}. By providing GPT with either part or a complete Verilog code $\mathcal{L}_{code}$ collected by~\cite{thakur2023benchmarking}, we inspire it to create a related Verilog design problem.

Process \circled{2} and \circled{3} help generate the initial design instruction pool $\mathcal{L}_{ins}$. After that, we iteratively augment this pool with mutation. Process \circled{4} applies two types of mutation operations on instructions sampled from the design instruction library $\mathcal{L}_{ins}$. The process \circled{5} would check every new design instruction using a set of rules and only passed valid instructions are added to $\mathcal{L}_{ins}$.

\textbf{Stage 3: Reference Code Generation.} The third stage targets generating the reference code corresponding to each instruction. As shown in \circled{6}, we feed each instruction from $\mathcal{L}_{ins}$ into GPT, generating corresponding reference design code. Then depending on whether process \circled{7} is applied, we generate two types of datasets in this flow, as we will introduce in Section~\ref{sec:RTLCoder-raw} and \ref{sec:RTLCoder-filter}, separately.




\subsection{80K Raw Dataset in RTLCoder-Data}
\label{sec:RTLCoder-raw}


To collect a large dataset, we continued to execute the basic data generation flow. Compared with the previous dataset from~\cite{liu2023rtlcoder}, we further enlarge the source code pool in $\mathcal{L}_{code}$ in process \circled{3} and continue the mutation in process \circled{4} during the generation process. In addition, we slightly relaxed the instruction checking conditions in process \circled{5}. Previously in~\cite{liu2023rtlcoder}, each new instruction is compared with all existing instructions in $\mathcal{L}_{ins}$ to check whether it introduces diversity. However, it takes a long time to compare each new instruction with all existing instructions. We removed this time-consuming diversity checking process, and only checked the basic instruction content in process \circled{5}. As we will introduce in Section~\ref{sec:data-evaluation}, results demonstrate that removing such diversity checking does not impair the overall diversity in the ultimate dataset. Finally, we accumulate and release a dataset of 80K samples, tripling the previous dataset in RTLCoder~\cite{liu2023rtlcoder}.

However, since the overall generation process of this 80K dataset relies on prompting commercial LLMs, we cannot guarantee the correctness of all samples. Therefore, we also refer it as a `raw' 80K dataset in RTLCoder-Data. To evaluate the effectiveness this `raw' dataset, we have trained LLMs with different numbers of data samples and evaluated results will be introduced in Section~\ref{sec:results}. Results indicate that a larger dataset leads to better model performance, and the performance is not saturated when 80K samples are used for training. Despite possible incorrectness in data samples, a larger dataset still clearly boosts model performance and proves useful.

\subsection{7K Verified Dataset in RTLCoder-Data}
\label{sec:RTLCoder-filter}


As introduced in Section~\ref{sec:RTLCoder-raw}, we have accumulated a raw dataset with 80K samples, but it is difficult to verify the correctness of each data sample. Specifically, it is feasible to automatically check the syntax correctness of the code in each sample with tools like VCS~\cite{vcs} or iVerilog, but it is very challenging to check whether the code has the correct functionality (i.e., code functionality matches the description in instruction). This functionality checking task is exactly hardware verification, which has been studied for decades, relies on human engineers, and is difficult to get guaranteed results. To the best of our knowledge, there is no prior work on automatic examination of code functionality correctness in dataset generation.

In this work, we made an innovative exploration to enable the automatic functionality checking of each instruction-code data sample. It is shown as the \emph{functionality checker} in process \circled{7} of Figure~\ref{fig:flow}. The solution is based on the LLM-assisted verification method introduced in Section~\ref{sec:asserteval}. First, based on the functionality description from instruction, we prompt commercial LLMs to generate corresponding assertions. The prompt techniques are from LLM-assisted verification works such as AssertLLM~\cite{fang2024assertllm}. Second, we combine the code and generated assertions, and feed them to verification platforms (e.g., JasperGold~\cite{Jasper}) to check whether the code violates any assertions. If all assertions are passed, it is likely that the code correctly implements the functionality described in the instruction. Still, this is not 100\% guarantee of sample correctness, but this process is fully automated and it leads to sufficiently high-quality samples for model training.

A problem in this verification-based functionality checking is, the assertions for verification are also generated by LLMs~\cite{fang2024assertllm}, thus the correctness of assertions is not guaranteed either. As a result, incorrect assertions may be generated for actually correct samples, making the correct samples fail the verification process. Therefore, this functionality checking method is conservative: Samples passing all assertions are likely to be correct, but correct data samples may fail the checking due to wrong assertions. Applying both functionality checking and syntax checking, as indicated by \circled{7}, we collected 7K high-quality verified samples. As we will introduce in Section~\ref{sec:results}, the 7K verified dataset leads to better LLM performance compared with models trained with even 50K raw data.

\section{RTL Generation Experiment Results} \label{sec:exper}

In this Section, we train and evaluate various LLM solutions with our 80K raw dataset and 7K verified datasets from RTLCoder-Data. In addition to extensive comparisons with various other LLM solutions, we studied the impact of training data amount, training scheme, and training data quality on LLM performance.

\begin{figure}[!b]
\includegraphics[width=0.49\textwidth]{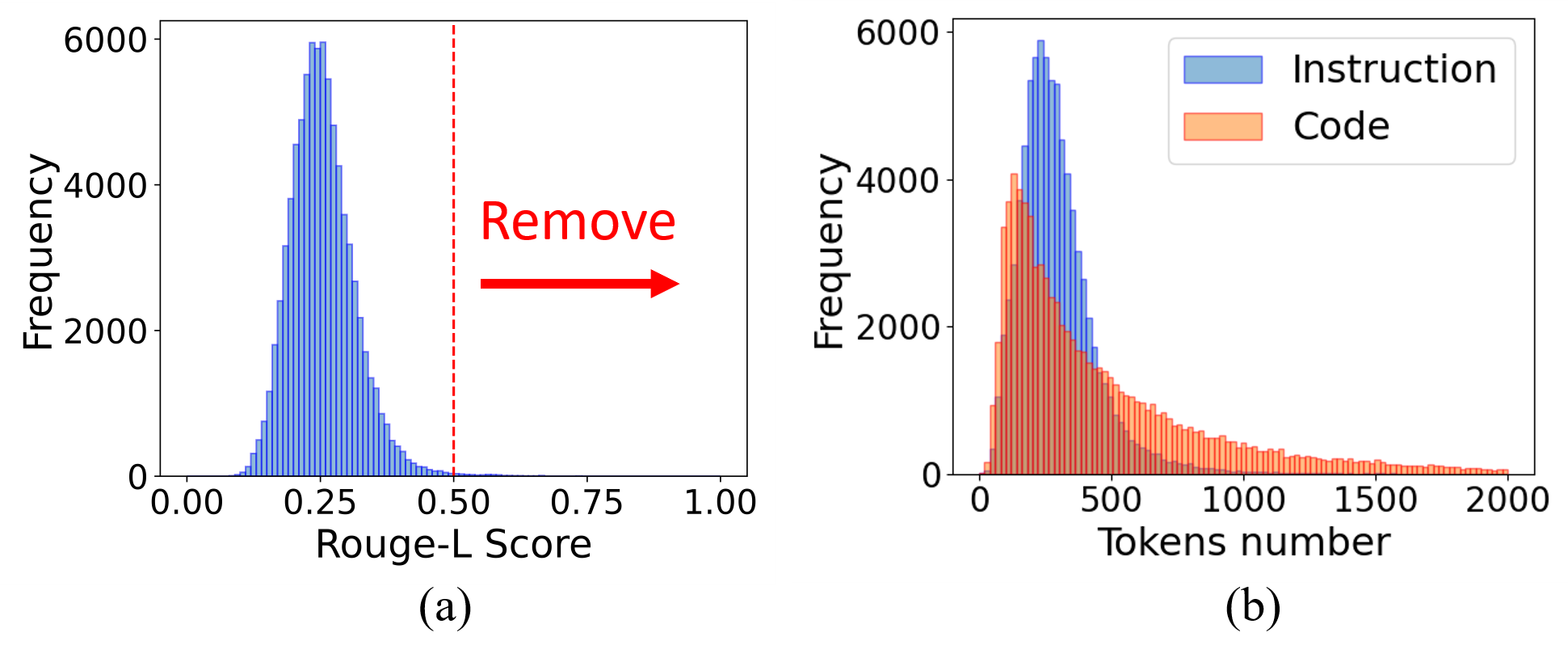}
\vspace{-.3in}
\caption{Training dataset analysis for the obtained 80K dataset. (a) Similarity measurement between training dataset and two benchmarks based on Rouge-L metric. (b) Tokens number distribution of instruction and code part.}
\vspace{-.1in}
\label{sigAnalysis}
\end{figure}

\subsection{LLM Training and Evaluation Setup}

To evaluate the performance of LLM-assisted RTL generation, we adopt two representative benchmarks named VerilogEval~\cite{liu2023verilogeval} and RTLLM~\cite{lu2023rtllm}. For RTLLM, following the original benchmark~\cite{lu2023rtllm}, each task is counted as success as long as \emph{any} of 5 trials passes the test. This can be interpreted as pass@5 metric. For all tested models, we evaluate all 3 $ temperature$ conditions $\left \{0.2, 0.5, 0.8\right \}$ and report the best performance for each model.

We choose the Mistral-7B-v0.1~\cite{jiang2023mistral} and DeepSeek-Coder-6.7b-Instruct~\cite{guo2024deepseek} as the basic pre-trained model for finetuning. In all experiments, we opted for the Adam optimizer with $\beta_1$ = 0.9, $\beta_2$ = 0.999, and learning rate $\gamma$ = 1e-5, while abstaining from the use of weight decay. Concurrently, we established a context length of 2048 and a global batch size of 256. We trained the model on only 4 consumer-level RTX 4090 GPUs (24GB each), each of which could afford $2\times2048$ context length using DeepSpeed stage-2~\cite{rasley2020deepspeed}.

\subsection{Evaluation of Dataset}
\label{sec:data-evaluation}


To prevent information leakage, for each instruction-code concatenated sample in the training dataset, we computed its maximum similarity with all test cases in the benchmarks. We employed the Rouge-L\footnote{The Rouge-L score $\in [0, 1]$, with values closer to 1 indicating higher similarity between the two sequences.}, a widely used similarity calculation metric in the LLM domain. As Figure ~\ref{sigAnalysis} (a) shows, most training samples have a low Rouge-L Value of around 0.25 and this indicates a low semantic overlap with the benchmarks. There are a small number of samples with higher similarity, and we get rid of these samples with Rouge-L > 0.5 during training. In addition,  Figure ~\ref{sigAnalysis} (b) shows that an instruction-code sample is generally within 2048 token length. So we can set 2048 as the max length in our finetuning.

To check the diversity of our proposed training dataset RTLCoder-Data-Raw (80K) and RTLCoder-Data-Verified (7K), we utilized two diversity measures: Compression Ratios (CR) and Part-of-Speech Compression Ratio (CR: POS) which are suggested best lexical diversity metrics by~\cite{shaib2024standardizing}.
CR is calculated utilizing text compression algorithms which can identify redundancy in the whole contents. The CR-POS can capture the repeated syntactic redundancy by compressing the part-of-speech (POS) tag sequences of the original text. We also followed the method utilized in ~\cite{shaib2024standardizing} to extract the tag sequences of the dataset.
The results are illustrated in Table~\ref{diversity}. Our proposed two datasets have lower CR and CR:POS than other existing open-source Verilog instruction-code datasets. This indicates that RTLCoder-Data-Raw (80K) and RTLCoder-Data-Verified (7K) have a satisfactory diversity.


\begin{table}[!t]
\centering
\resizebox{0.49\textwidth}{!}{
\begin{tabular}{|c||c|c|c|c|}
\hline
        & \begin{tabular}[c]{@{}c@{}}RTLCoder-Data\\ Raw (80K)\end{tabular} & \begin{tabular}[c]{@{}c@{}}RTLCoder-Data\\ Verified (7K)\end{tabular} & \begin{tabular}[c]{@{}c@{}}MG-Verilog\\ ~\cite{zhang2024mg}\end{tabular} & \begin{tabular}[c]{@{}c@{}}Goh et al.\\\cite{goh2024english}\end{tabular} \\ \hline\hline
CR      & 4.21                                                          & 4.32                                                                  & 5.80       & 5.27                                                           \\ \hline
CR: POS & 7.33                                                          & 7.45                                                                  & 9.16       & 10.1                                                           \\ \hline
\end{tabular}
}
\vspace{.1in}
\caption{Diversity scores (CR, CR:POS) of RTLCoder-Data Raw (80K), RTLCoder-Data Verified (7K), and other RTL datasets~\cite{zhang2024mg, goh2024english}. Lower CR and CR:POS mean higher dataset diversity. Both datasets from RTLCoder-Data exhibit satisfactory diversity compared with others.}
\label{diversity}
\vspace{-.25in}
\end{table}

\begin{table*}[!t]
\centering
\vspace{-.05in}
\resizebox{0.98\textwidth}{!}{
\renewcommand{\arraystretch}{1.1}
\begin{tabular}{|c|c|c||c|c|c||c|c|c||c|c|} 
\hline
\multirow{4}{*}{Model Type}  & \multirow{4}{*}{Evaluated Model} & \multirow{3}{*}{Num of} & \multicolumn{6}{c||}{VerilogEval Benchmark~\cite{liu2023verilogeval}} & \multicolumn{2}{c|}{RTLLM V1.1~\cite{lu2023rtllm}}  \\ 
   &   &  \multirow{3}{*}{Params}  & \multicolumn{6}{c||}{(using pass@k metric)}  &  \multicolumn{2}{c|}{(using pass@5 metric)}  \\
\cline{4-9}\cline{10-11}
   &     &     & \multicolumn{3}{c||}{Eval-Machine (\%)} & \multicolumn{3}{c||}{Eval-Human (\%)}  &  \multirow{2}{*}{Syntax-VCS(\%)}   &  \multirow{2}{*}{Func (\%)}   \\ 
\cline{4-9} 
   &     &   &  k=1 &  k=5 &  k=10  &  k=1 &  k=5 &  k=10  &      &    \\
\hline
\hline
\multirow{4}{*}{Closed-Source}      & GPT-3.5     & N/A       & 46.7  &69.1  &74.1       & 26.7   &45.8  &51.7                &89.7    &37.9     \\
\multirow{4}{*}{Baseline}           & GPT4        & N/A       &   60.0  &70.6  &73.5       & \cellcolor{lightgreen}43.5      &\cellcolor{lightgreen}55.8  &\cellcolor{lightgreen}58.9                   &\cellcolor{lightgreen}100    &\cellcolor{lightgreen}65.5      \\
                        &   ChipNeMo~\cite{liu2023chipnemo}  &13B &43.4  &N/A &N/A &22.4  &N/A &N/A   & N/A & N/A              \\
                        & VerilogEval~\cite{liu2023verilogeval}     & 16B     & 46.2   &67.3  &73.7      & 28.8   &45.9  &52.3   & N/A    & N/A        \\ 
                        & BetterV~\cite{pei2024betterv}  & 7B  &  \cellcolor[HTML]{C5D9F1} 
 64.2 & 75.4   &   79.1        &  40.9 &  50.0 &53.3
                          &N/A &N/A    \\ 
\hline
\hline
\multirow{4}{*}{Open-Source}            & Codegen2~\cite{nijkamp2023codegen2}  & 16B  & 5.00  &9.00  &13.9        & 0.90   &4.10  &7.25     &72.4 &6.90                           \\
\multirow{4}{*}{Baseline}               & Starcoder~\cite{li2023starcoder}    & 15B  & 46.8  &54.5  &59.6   & 18.1    &26.1  &30.4  &\cellcolor{lightred}93.1 &27.6                          \\
                                    & Thakur et al.~\cite{thakur2023benchmarking}   & 16B    & 44.0  &52.6  &59.2       & 30.3  &43.9  &49.6     &86.2    &24.1        \\ 
& Mistral-7B~\cite{jiang2023mistral}       & 7B     & 36.9  &48.8  &57.4            &4.49    &12.6  &18.6        &72.4    &20.7       \\
 &  DeepSeek-Coder~\cite{guo2024deepseek}        & 6.7B     & 54.1  &63.8  &67.5            &30.2    &42.2  &46.2      &89.6    &34.5         \\
 \hline
\hline
\textbf{Scoring-based}
               & Mistral-Scoring (27K data samples)      & 7B  &  62.5    &72.2  &76.6        & 36.7   &45.5 &49.2            &\cellcolor[HTML]{C5D9F1}96.6 &\cellcolor{lightred}48.3    \\
\textbf{Training~\cite{liu2023rtlcoder}}  
                &  DeepSeek-Scoring (27K data samples)      & 6.7B  &  61.2    &  \cellcolor[HTML]{C5D9F1}76.5  &\cellcolor{lightgreen}81.8         & \cellcolor{lightred}{41.6}   &\cellcolor{lightred}50.1 &    53.4               &\cellcolor{lightred}93.1 &\cellcolor{lightred}48.3                        \\
\hline
\hline
\multirow{4}{*}{\textbf{Basic Direct}}  
         &  Mistral-Direct (27K data samples)          & 7B            & 58.9   &70.0 &74.1            &34.4     &42.3  &45.1      &89.7    &41.4         \\ 
        &  DeepSeek-Direct (5K data samples)          & 6.7B         &53.7    &71.7  &77.1        & 32.9    &45.8  &52.4     &\cellcolor{lightred}93.1     &41.4       \\ 
\multirow{2}{*}{\textbf{Training}} 
        &  DeepSeek-Direct (27K data samples)        & 6.7B            &59.8    &       73.6  &     77.2            &39.1       &48.3                   &51.3        &86.2    &44.8   \\
       &  DeepSeek-Direct (50K data samples)         & 6.7B         &  \cellcolor{lightred}62.6    &  75.6  &   \cellcolor{lightred}80.5    &  38.9     &48.7  &51.8     & 89.7    &\cellcolor[HTML]{C5D9F1}55.2   \\
        &  DeepSeek-Direct (80K data samples)         & 6.7B         &  \cellcolor{lightgreen}64.7    & \cellcolor{lightgreen}76.6     &   \cellcolor[HTML]{C5D9F1}80.8            &  \cellcolor[HTML]{C5D9F1}42.8     &   \cellcolor[HTML]{C5D9F1}51.6  &  \cellcolor{lightred}55.0     &\cellcolor{lightred}93.1    & \cellcolor{lightred}48.3  \\
\hline
\hline
\textbf{Verified Dataset}
               & DeepSeek-Direct (7K verified data samples)  & 7B  &  61.3  &  \cellcolor{lightred}76.3  & \cellcolor[HTML]{C5D9F1}80.8       & 38.9   & \cellcolor{lightred}50.1 &    \cellcolor[HTML]{C5D9F1}55.3            &\cellcolor{lightgreen}100 &\cellcolor{lightred}48.3 
\\
\hline
\end{tabular}
}
\vspace{.08in}
\caption{Performance comparison of RTL code generators on VerilogEval Benchmark~\cite{liu2023verilogeval} and RTLLM Benchmark~\cite{lu2023rtllm}. The top scores ranked 1$^{\text{st}}$, 2$^{\text{nd}}$, and 3$^{\text{rd}}$ in each column are marked in \colorbox{lightgreen}{Green}, \colorbox[HTML]{C5D9F1}{Blue}, and \colorbox{lightred}{Red}, respectively.} 
\label{tbl:result}
\vspace{-.15in}
\end{table*}

\begin{figure}[!t]
\includegraphics[width=0.47\textwidth]{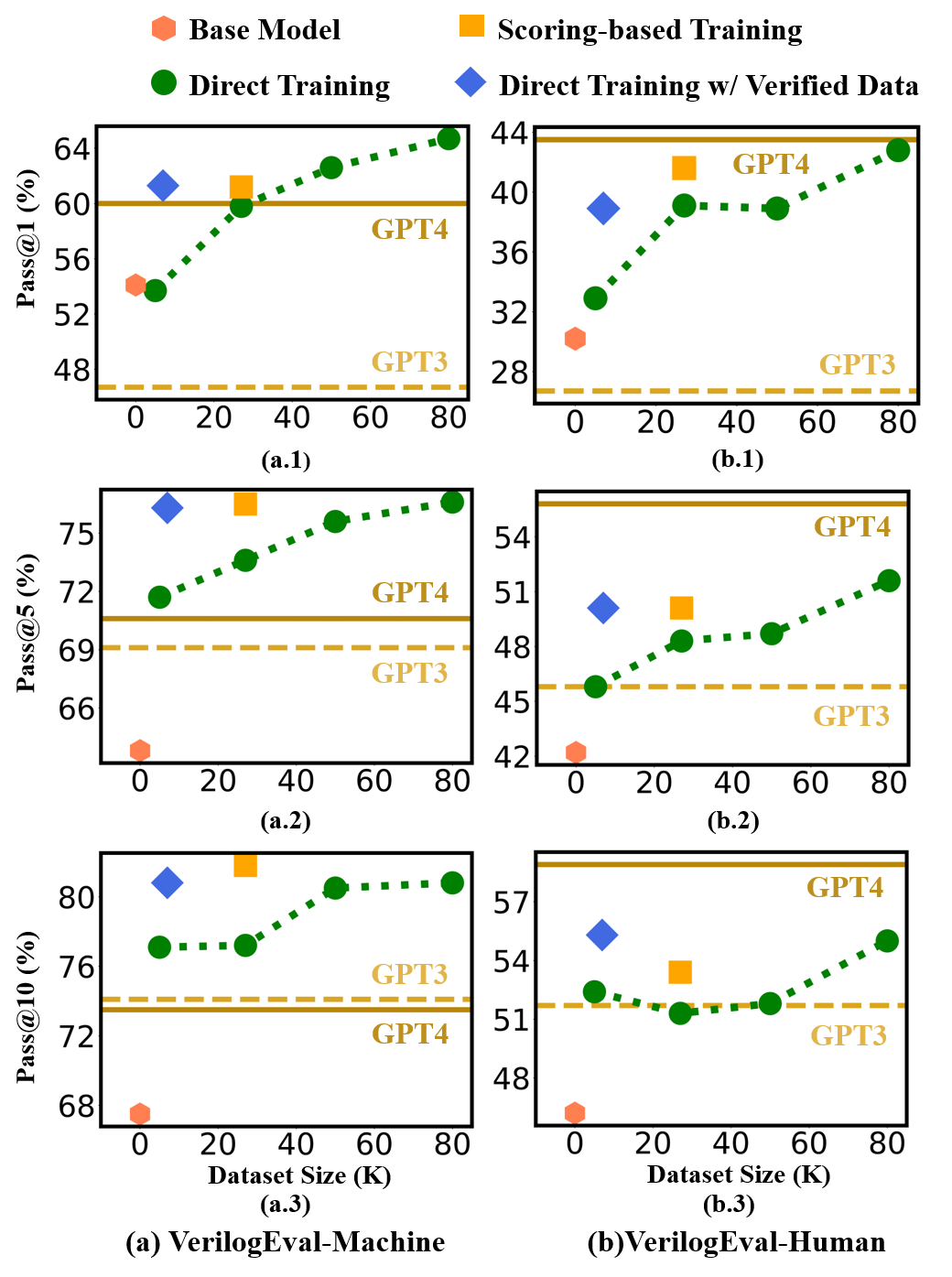}
\vspace{-.2in}
\caption{The pass@k performance on VerilogEval benchmarks versus the amount of training data from RTLCoder-Data. The performance improves as the data size increases.} 
\vspace{-.15in}
\label{adddata}
\end{figure}

\subsection{Result of Trained LLM in RTL Generation}
\label{sec:results}

Table~\ref{tbl:result} summarizes the comparison of various LLM-assisted RTL generation solutions, including commercial models GPT3.5/GPT4, both closed- and open-source LLMs customized for Verilog generation~\cite{liu2023verilogeval, thakur2023benchmarking, pei2024betterv}, general software code generators~\cite{nijkamp2023codegen2, li2023starcoder, jiang2023mistral}, and our fine-tuned models based on DeepSeek-Coder-6.7b-Instruct~\cite{guo2024deepseek} and Mistral-7B-v0.1~\cite{jiang2023mistral} with different amount of training data and training schemes. Relevant results are also presented in Figure~\ref{adddata}, which shows LLM performance versus the amount of training data.


\textbf{Overall performance based on RTLCoder-Data.} We train the base model directly on the RTLCoder-Data Raw (80K) through instruction-supervised fine-tuning which is referred to as ``basic direct training" in Table~\ref{tbl:result}. We can observe that DeepSeek-Direct (80K data samples) outperforms all other baseline models in Eval-Machine and is only inferior to GPT-4 in Eval-Human and RTLLM V1.1. Specifically, in the Eval-Machine part, it even outperforms GPT4 by an absolute value of 4.7\% in the pass@1 metric. In summary, DeepSeek-Direct (80K data samples) outperforms GPT-3.5 and all non-commercial baselines in all metrics. It is surprising that the lightweight model with only 7 billion parameters could achieve such impressive accuracy despite its smaller size.


\textbf{Impact of training data amount.} To further investigate the impact of dataset size on model performance, we sampled subsets of 5K, 27K, and 50K samples from the RTLCoder-Data Raw (80K) and then conducted direct finetuning on these subsets. The results are shown in Table~\ref{tbl:result} and also plotted in Figure~\ref{adddata}. We can observe that as the training data volume increases, the overall performance of the model on the benchmarks also improves. For instance, as the training data size increases from 5K to 80K, the model's performance on the Eval-Machine pass@1 metric rises from 53.7\% to 64.7\%. Additionally, as illustrated in Figure~\ref{adddata}, even with 80K data samples, there are still no signs of model performance saturation. This indicates that enlarging the training dataset size can significantly boost the model's code generation capabilities.

\textbf{Impact of training scheme.}  We extracted a 27K subset from the RTLCoder-Data Raw (80K) and employed the code quality feedback-based training scheme proposed in RTLCoder~\cite{liu2023rtlcoder} to obtain models named Mistral-Scoring and DeepSeek-Scoring. Their performance is presented in Table~\ref{tbl:result} and Figure~\ref{adddata} titled `Scoring-based Training'. Compared with DeepSeek-Direct (27K data samples) and Mistral-Direct (27K data samples), models trained with the scoring-based scheme are better on all benchmarks, indicating that this better scoring-based training method~\cite{liu2023rtlcoder} improves model performance.

\textbf{Impact of training data quality.}  In the `Verified Dataset' part of Table~\ref{tbl:result}, we directly trained the DeepSeek-Coder using the dataset RTLCoder-Data Verified (7K). DeepSeek-Direct (7K verified) outperforms DeepSeek-Direct (27K) across all benchmarks and even surpasses DeepSeek-Direct (50K) on 6/8 metrics. Moreover, DeepSeek-Direct (7K verified) only uses $< 20\%$ of the training time of DeepSeek-Direct (50K). This demonstrates that enhancing the quality of the training dataset can improve the model performance and reduce the LLM training cost. It indicates the great potential of our proposed assertion-based functionality checking technique.







\section{Limitation and Challenges}

Finally, we would like to discuss some challenges and questions we encountered during the development of the dataset or benchmark for LLM-assisted design automation solutions, and share our thoughts about these questions.

When building the open-source benchmark for RTL generation, we encountered several challenges:
\begin{enumerate}
    \item \textit{Shall we include more complex designs in the benchmark?} Due to the limited abilities of existing LLMs, almost all LLMs encounter difficulty in generating `correct' RTL design code for very complex designs. As a result, overly complex designs often fail to differentiate the capabilities of the models. In addition, it is difficult to precisely describe complex designs with natural languages. 
    \item \textit{How detailed should the description be?} 
    When descriptions are overly vague or general, LLMs struggle to produce designs that meet expected functionality, making it difficult to assess model capabilities. Conversely, if descriptions are too detailed, focusing on intricate RTL circuit specifics, the RTL generation effectively becomes a form of `code translation', which also fails to demonstrate the general generative abilities of LLMs. Therefore, the level of detail in the description for benchmarking requires careful consideration.  
     \item \textit{How to alleviate the influence of training data leakage on the benchmark scores?} The overlap between the training dataset and benchmarks should always be carefully examined because an overfitted LLM cannot generalize well in practice. Overfitted LLM can easily lead to unfair comparisons and misleading conclusions. However, the text similarity approximation we used based on Rouge-L metric may not be perfect. In addition, leakage during the LLM pre-training process is difficult to control. How to define and evaluate the data leakage in RTL generation is still a challenging open problem.
\end{enumerate}


The main challenge in LLM-based assertion generation centers around improving the quality of the generated assertions. We break down this challenge into two key questions:
\begin{enumerate}
     \item \textit{How to better quantify the assertion quality? } Existing metrics like syntax/semantics correctness and COI coverage are useful but inadequate for complex verification scenarios, such as capturing state transitions or ensuring different assertions cover distinct properties. More precise evaluation techniques are worth exploring in future works.
    \item \textit{What limits the generation of high-quality assertions?} High-quality assertions depend not only on LLM capabilities but also on the richness of the specification documents. Specifications that lack detailed functionalities or connectivities will limit the effectiveness of assertion generation, regardless of the capability of LLM.
\end{enumerate}

\section{Conclusion}\label{sec:concl}

In this work, we present our latest advances in open-source benchmarks and datasets for developing LLMs to assist in design RTL generation and verification. We fully open-sourced 1) RTLLM 2.0, an updated benchmark for the evaluation of LLM-assisted RTL generation; 2) AssertEval, a benchmark or the evaluation of LLM-assisted assertion generation for verification; and 3) RTLCoder-Data, an extended open-source dataset for training LLMs for RTL generation. It provides 80K instruction-code data samples, as well as a 7K verified high-quality dataset. These open-source circuit data are provided as off-the-shelf resources, targeting more democratized and reproducible AI for EDA research.





\clearpage
\bibliographystyle{ACM-Reference-Format}
\bibliography{ref, assert}

\end{document}